\documentstyle[12pt]{article}
\newcommand{\beq}{\begin{equation}}
\newcommand{\eeq}{\end{equation}}
\newcommand{\id}
 {i\kern.06em\hbox{\raise.25ex\hbox{$/$}\kern-.60em$\partial$}}

\newcommand{\bs}{/\kern-.52em b}
\newcommand{\qs}{/\kern-.52em s}
\newcommand{\D}{{\cal{D}}}
\newcommand{\dv}{\!d^3\!x\,}

\newcommand{\dd}
{\kern.06em\hbox{\raise.25ex\hbox{$/$}\kern-.60em$\partial$}}

\newcommand{\tr}{\mathop{\rm tr}\nolimits}
\begin{document}

\title{Non Abelian Bosonization in Two and Three Dimensions}
\author{
J.C. Le Guillou$^a$\thanks{Also at {\it Universit\'e de Savoie} and at
{\it Institut Universitaire de France}}\,,
E. Moreno$^b$, C. N\'u\~nez$^c$ \\
and\\
F.A. Schaposnik$^c$\thanks{Investigador CICBA, Argentina}
\\
~
\\
{\normalsize\it
$^a$Laboratoire de Physique Th\'eorique ENSLAPP}
\thanks{URA 1436 du CNRS associ\'ee
\`a l'Ecole Normale Sup\'erieure de Lyon et \`a
l'Universit\'e de Savoie}\\
{\normalsize\it
LAPP, B.P. 110, F-74941 Annecy-le-Vieux Cedex, France}\\
~\\
{\normalsize\it
$^b$ Physics Department, City College of the City University
of New York}\\
{\normalsize\it
New York NY 10031, USA}
 \\
~\\
{\normalsize\it
$^c$Departamento de F\'\i sica, Universidad Nacional de La Plata}\\
{\normalsize\it
C.C. 67, 1900 La Plata, Argentina}}
\date{}
\maketitle

\vspace{-6 in}
\hfill\vbox{
\hbox{~~}
\hbox{\it Modified version}
\hbox{\today}
}
\vspace{5 in}

\begin{abstract}
{We discuss  non-Abelian bosonization of two and three dimensional
fermions using a  path-integral framework in  which the
bosonic action follows from the evaluation of the
fermion determinant for the Dirac operator in the
presence of a vector field. This naturally leads to
the Wess-Zumino-Witten action for massless
two-dimensional fermions and to
a Chern-Simons action for very massive three dimensional
fermions. One advantage of our approach is
that it allows to derive the exact bosonization
recipe for fermion currents in a systematic way.   }
\end{abstract}
\newpage
\section{Introduction and results}
In this paper we investigate the problem of finding the
mapping of a three-dimensional free fermion theory with non-abelian
symmetry onto an equivalent bosonic quantum field theory. This
mapping, commonly called bosonization, has been already
established along the lines of the present investigation
for the case of abelian symmetry \cite{FS}-\cite{LNS} and
it has been also discussed in the non-abelian case,
using a related method, in \cite{B}. In all these
investigations we employ an approach
close to that put forward in
a series of very interesting works on smooth
bosonization and
duality  bosonization
\cite{DNS1}-\cite{DNS5}. Other related or alternative
approaches to
bosonization in $d>2$ dimensions have been also developed
\cite{TSSG}-\cite{BBG}.

The advantage of the bosonization method
that we employ here lies on the fact that it provides
a systematic procedure for deriving $d \ge 2$ bosonization recipes both
for abelian and non-abelian symmetries. In this way, it gives an
adequate framework for obtaining the bosonic equivalent of the original
fermionic theory, the recipe for mapping fermionic
and bosonic currents  as well as the current commutation relations
which are at the basis of bosonization.

The approach we follow starts from  the path-integral
defining the generating functional
for a theory of free fermions (including sources
for fermionic currents) and ends with
the generating functional for an equivalent
bosonic theory. This allows to identify, {\it exactly},
the bosonization recipe for fermion currents
independently of the number of space-time dimensions.
Interestingly enough,
one follows a series of steps which
are the same for any space-time dimension and both
for abelian and non-abelian  symmetries.
Of course, apart from the two-dimensional
case and from the
large fermion mass limit in
the three dimensional case, 
one only achieves in general a partial bosonization in the sense
that one cannot compute exactly the fermion path-integral in order to
derive a local bosonic Lagrangian.

Extending the three dimensional
abelian bosonization approach discused
in \cite{FS}-\cite{LNS}
 we derive
in the present paper 
three dimensional
non-abelian bo\-so\-niza\-tion. Concerning
the fermion current, our method
allows to derive the exact bosonization recipe
\beq
\bar \psi^i \gamma_\mu t^a_{ij} \psi^j \to
\pm \frac{i}{8\pi}\varepsilon_{\mu \nu \alpha}
\partial_\nu A^a_\alpha \,\, ,
\label{Ssss}
\eeq
where $i,j=1,\cdots,N$, $a=1,\cdots,{\rm dim}G$,
$t^a$ are the generators and $f^{a b c}$  the
structure constants of the symmetry group G.
Finally $A_\mu$ is a vector field taking values in the
Lie algebra of $G$.

The knowledge of the bosonic action accompanying
this bosonization rule is necessarily
approximate since it implies  the evaluation
of the $d=3$  determinant for fermions
coupled to a vector field. We then consider 
the case 
of very massive free fermions
 showing  that, within this approximation,
the fermion lagrangian bosonizes to 
 the non-abelian Chern-Simons term,
\beq
 \bar\psi^i \left(i\gamma^{\mu} \partial_{\mu}+m \right) \psi^i  \to
\mp \frac{i}{8 \pi} \epsilon_{\mu \nu\alpha} \left(
A_{\mu}^a \partial_{\nu}A^a_{\alpha} + 1/3 f^{a b c}
A_{\mu}^a A_{\nu}^b A_{\alpha}^c \right) ~,
\label{aq11}
\eeq
This result, advanced in \cite{B} using a completely
different approach, is the natural extension of the
abelian result \cite{FS}-\cite{LNS}. In this last
case the bosonic theory   corresponds, in the large fermion mass
limit, to a Chern-Simons theory while in the massless case
it coincides with the abelian
non-local action discussed in \cite{Mar},\cite{BFO}. In
fact one should expect that an analysis similar to that
in \cite{BFO} can be carried out leading to an
explicit (although complicated) bosonic action valid in all
fermion mass regimes.

The results above are derived in section 3. As a warming up
exercise, we rederive in section 2 the bosonization rules for
two-dimensional non-abelian models. Indeed, following
an approach related to
that developed in \cite{BQn} for two-dimensional
bosonization we arrive
 to the Wess-Zumino-Witten action
and  the well-known bosonization recipe for
fermion currents
\beq
j_+ \to \frac{i}{4\pi}a^{-1} \partial_+ a
\label{ele1}
\eeq
\beq
j_-  \to \frac{i}{4\pi}a \partial_- a^{-1}
\label{ele2}
\eeq
with $a$  the group-valued bosonic field with
dynamics governed by the Wess-Zumino-Witten
action.
Now, these rules, as well as the Polyakov-Wiegmann identity
that we repeatedly use in its  derivation,
deeply relay on the holomorphic properties of
two-dimensional theories \cite{PWI}\cite{Wi0}. Strikingly,
we found that in
the three dimensional case, a BRST symmetry structure underlying  the
bosonic version of the fermionic generating functional
plays a similar role and allows to end, at least
in the large mass limit, with a simple bosonization rule.
This BRST symmetry is highly related to that used in 
\cite{DNS1}-\cite{DNS3}, \cite{DNS4}-\cite{TSSG},
and is 
analogous to that arising
in topological field theories  \cite{BS}-\cite{BRT},
its origin being related to the
way the originally ``trivial''  bosonic field enters into play.

\section{Warming up: $d=2$ non-abelian bosonization}

Non-abelian bosonization in two dimensional space time was formulated
first by Witten \cite{Wi} by comparing the current algebra for free fermions
and for a bosonic sigma model with a Wess-Zumino term. Afterwards,
different approaches rederived and discussed
the bosonization recipe \cite{dVR}-\cite{GR} but in
general they were not constructive in the sense that the bosonic
theory was not obtained from the fermionic one by following a
series of steps that could be generalized to other cases, in particular
to a possible higher-dimensional bosonization. More recently, using the
the duality technique \cite{BQ}-\cite{BLQ} which has as 
starting point the
smooth bosonization approach \cite{DNS1}-\cite{DNS3}, 
\cite{DNS4}-\cite{DNS5}
the recipe for
non-abelian bosonization was obtained by Burgess and Quevedo \cite{BQn}
in a way which is more adaptable to generalizations to higher dimensions.
The approach to two-dimensional bosonization that we present in this section
is  related to that in \cite{BQn} and we think  that it
is worthwhile to describe it here in detail since it provides many of the
clues which allow us  to derive the
non-abelian bosonization recipe for 3-dimensional fermions.

Following Witten  \cite{Wi}
one can see that the bosonic picture for a theory of
$N$ free massless Dirac fermions corresponds
to a bosonic field $a \in SU(N)$
with a Wess-Zumino-Witten action and a real bosonic field $\phi$
with a free scalar field action
\cite{dVR}-\cite{GO}. Since we shall be mainly interested
in the specific non-abelian aspect of bosonization, we will not discuss
the $\phi$ sector of the corresponding bosonic theory (although
the method can trivially take into account the $U(1)$ sector
associated with it).

We start from the (Euclidean) Lagrangian for free massless Dirac fermions
in $2$ dimensions
\beq
L = \bar\psi (\id) \psi
\label{L}
\eeq
where fermions are in the fundamental representation of some group $G$.
The corresponding
generating functional reads
\beq
Z_{fer}[s]
= \int D\bar \psi D\psi
\exp[-\int d^2x  \bar\psi (\id + \qs) \psi]
\label{Zs}
\eeq
with $s_\mu = s_\mu^a t^a$  an external source
taking values in the Lie algebra of
$SU(N)$.

Our derivation of bosonization rules both in $2$ and $3$ dimensions
heavily relies on the invariance of the measure under local transformations
of the fermion variables, 
$ \psi \to h(x)  \psi$,  $\bar \psi \to \bar \psi  h(x)^{-1}$ with $h \in G$.
As a consequence
the generating functional (\ref{Zs}) is automatically invariant under
local transformations of the source
\beq
s_\mu \to s_\mu^h = h^{-1} s_\mu h + i h^{-1} \partial_\mu h
\label{inva}
\eeq
\beq
Z[s^h] = Z[s]
\label{idiss}
\eeq
In view of this, if we perform the change of variables
\[
\psi =g(x) \psi'
\]
\beq
\bar \psi = \bar \psi' g^{-1}(x) .
\label{chas22}
\eeq
$Z_{fer}[s]$ becomes
\beq
Z_{fer}[s] =  \int \D \bar \psi \D \psi \D g
\exp[-\int d^2x \bar \psi (\id + \qs ^g ) \psi]
\label{n2s}
\eeq
where an integration over $g$ has been included since it just amounts
to a change of normalization. Integrating out fermions we have
\beq
Z_{fer}[s] = \int \D g \;\det(\id +  {\qs}^g)
\label{cambios}
\eeq
Now, posing
\beq
{s}_\mu^g =  b_\mu
\label{nuezz}
\eeq
and using
\beq
f_{\mu \nu}[b] = g^{-1} f_{\mu \nu}[s] g
\label{rela}
\eeq
we shall trade the $g$ integration for an integration over
connections $b$ satisfying condition (\ref{rela}).
To this end we shall use the $d=2$ identity (proven in the Appendix)
\beq
 \int \D b_\mu {\cal H}[b]
\delta[\varepsilon_{\mu \nu} (f_{\mu\nu}[b] - f_{\mu\nu}[s])]
=  \int \D g {\cal H}[s^g]
\label{idle}
\eeq
Here ${\cal H}$ is a gauge invariant function. Identity (\ref{idle})
 allows us to write eq.(\ref{cambios}) in the form
\beq
Z_{fer} =  \int \D b_\mu \Delta \delta(b_+ - s_+)
\delta\left[\varepsilon_{\mu\nu}(f_{\mu \nu}[b]- f_{\mu \nu}[s])\right]
\det(\id + \bs) \, .
\label{n3ss}
\eeq
For convenience, we have chosen to fix the gauge using the
condition $b_+ = s_+$  being $\Delta$ the corresponding
Faddeev-Popov determinant.

We
now introduce a Lagrange multiplier ${\hat a}$
(taking values in the Lie algebra of $G$) to enforce the
delta function condition
\begin{eqnarray}
Z_{fer}[s]  & = &
\int \D {\hat a}\D b_\mu \,\Delta \delta(b_+ - s_+)
\det[\id +\bs]  \times \nonumber \\
& &
\exp \left(-\frac{C}{8\pi}
 tr \int d^2x  {\hat a} \,\varepsilon _{\mu \nu }
(f_{\mu \nu }[b]
 - f_{\mu \nu }[s])\right)
\label{copia}
\end{eqnarray}
with $C$ a constant to be conveniently adjusted.
We now write sources and the $b_\mu$ field in terms of group-valued
variables,
\beq
s_+ = i {\tilde s}^{-1} \partial_+ \tilde s
\label{s+}
\eeq
\beq
s_- = i s \partial_- s^{-1}
\label{s-}
\eeq
\beq
b_+ = i (\tilde b \tilde s)^{-1} \partial_+ (\tilde b \tilde s)
\label{b+}
\eeq
\beq
b_- = i (sb) \partial_- (sb)^{-1}
\label{ga}
\eeq
so that the fermion determinant can be related to the
Wess-Zumino-Witten action \cite{PWI},
\beq
\det[\id +\bs] = \exp (W[\tilde b \tilde s s b])
\label{deter}
\eeq
In writing eq.(\ref{deter}) a gauge-invariant regularization
is assumed so that the left and right-handed sectors
enter in gauge invariant
combinations.
In this way, gauge transformations of
the source $s_\mu$, which as stated before,
should leave the generating functional invariant,
do not change the determinant. This will be
the criterion we shall adopt each time
determinants (always needing a regularization)
has to be computed.
Concerning the Jacobian for passing from the $b_\mu$ variable to the
$b, \tilde b$ one can easily show that
\beq
 \int \Delta \delta(b_+ - s_+)\D b_\mu = \int \exp(\kappa W[\tilde b
\tilde s s b])
 \delta( \tilde b - I)  \D \tilde b \D b
\label{jaci}
\eeq
so that $Z_{fer}[s]$ becomes
\begin{eqnarray}
Z_{fer}[s] & = &
\int \D {\hat a} \D b \exp \left(i\frac{C}{4\pi}
tr \int d^2x
( D_+[\tilde  s] {\hat a}) \, s b ( \partial_- b^{-1}) s^{-1}\right)
\times \nonumber\\
& &
 \exp \left((1+\kappa)W[ \tilde s s b] \right)  \, .
\label{copiag}
\end{eqnarray}

A convenient change of variables to pass from
integration over the algebra valued
 Lagrange multiplier ${\hat a}$ to
a group valued variable $a$ is the one defined through
\beq
D_+[\tilde s] {\hat a} = {\tilde s}^{-1} (a^{-1}\partial_+a) \tilde s
\label{nuez}
\eeq
Calculation of the
corresponding jacobian $J_{L} $
\beq
\D \hat a = J_{L} \D a
\eeq
should be carefully done. Indeed, in
the present approach
to bosonization, it is the group valued variable $a$ who plays
the role of the boson field equivalent to the original fermion field.
Since for the latter (a free fermi field) there was no local symmetry,
the former should not be endowed with this symmetry.
With this in mind, we shall
maintain $a$ unchanged under local transformations $g$
while transforming  $\hat a$,
$\hat a \to g^{-1} \hat a g$,
so that eq.(\ref{nuez})  changes covariantly when
one simultaneously changes  $\tilde s \to \tilde s g$.
One can easily prove that
\beq
J_L = \exp (\kappa W[a \tilde s s] - \kappa W[\tilde s s])
\label{nuezi}
\eeq
so that the generating functional reads
\begin{eqnarray}
 Z_{fer}[s] & =
\int \D a \D b \exp((1 +\kappa)W[\tilde s s b]) \times
\exp \left(\kappa (W[a \tilde s s]- W[\tilde s s])\right)
\times \nonumber\\
 &\exp(-\frac{C}{4\pi} tr \int d^2x {\tilde s}^{-1}
(a^{-1} \partial_+ a) \tilde s s (b \partial_- b^{-1}) s^{-1} ) \, .
\label{copg}
\end{eqnarray}
If one repeatedly uses Polyakov-Wiegmann identity
and chooses the up to now arbitrary
constant $C$ as
\beq
-C = 1 + \kappa
\label{D}
\eeq
one can write $Z_{fer}[s]$ in the form 
\begin{equation}
 Z_{fer}[s] =   \int \D a  \D b \;
\exp (W[\tilde s s] - W[ a \tilde s s])
\exp\left((1+\kappa)W{[a \tilde s s b ]} \right)
\label{fifi}
\end{equation}
Now, the $b$ integration can be
 trivially factorized this leading to 
\beq
Z_{fer}[s] =
\int \D a \;\exp(-W[a \tilde s s] + W[\tilde s s] ) \,.
\label{su}
\eeq
or, after the shift
$ a {\tilde s} s \to \tilde s a s$
\label{ayy}
\begin{eqnarray}
Z_{fer}[s_+]&  =  &
\int \D a \;\exp(-W[a] + \frac{i}{4\pi} tr \int d^2x (s_+ a \partial_- a^{-1}
+ \nonumber \\
& & s_- a^{-1} \partial_+ a) ) \times
\exp(\frac{1}{4\pi} tr \int d^2x (a^{-1} s_+ a s_- -  s_+ s_-))
\label{chuchi}
\end{eqnarray}
We have then arrived to the identity
\beq
Z_{fer}[s] = Z_{bos}[s]
\label{cg}
\eeq
where $Z_{bos}[s]$ is
the generating function for a Wess-Zumino-Witten model. Differentiation with
respect to any one of the
two sources gives correlation functions in a given chirality sector.
The answer corresponds to Witten's  bosonization recipe \cite{Wi}
\beq
\bar \psi t^a \gamma_+ \psi \to \frac{i}{4\pi} a^{-1} \partial_+ a
\label{rec1}
\eeq
\beq
\bar \psi t^a \gamma_- \psi \to  \frac{i}{4\pi} a \partial_- a^{-1} \, ,
\label{rec2}
\eeq
the l.h.s. to be computed in a free fermionic model, the r.h.s. in
a Wess-Zumino-Witten model.


\section{ $d=3$ non-abelian bosonization}
Contrary to the case of  two-dimensional massless
fermions,
one cannot compute exactly the   Dirac operator determinant
for $d > 2$
in the presence of an arbitrary gauge field $b_\mu$,
neither in the massless nor in the massive case. This
implies the necessity of making  approximations at some stage of
our bosonization  procedure to render calculations feasible.
In the $d=3$ Abelian case
one can handle these approximations in a very general framework
\cite{BFO},\cite{LNS}. Being the non-Abelian case far more complicated
than the Abelian one,  we shall only discuss the limiting case
 of very massive fermions, for which
 the fermion determinant is related to the Chern-Simons (CS)
action \cite{NS}-\cite{Red}.

A second problem arising when one tries to extend  the non-Abelian
boson-fermion mapping from $d=2$ to $d=3$
concerns the central role that plays the Polyakov-Wiegman identity,
related to the holomorphic
properties of the two-dimensional model \cite{Wit}. In principle, a
$3$-dimensional analogue of this identity is not available and
this forbids a trivial extension to $d=3$  of the procedure
described in the precedent section for two-dimensional
bosonization. However, as we shall see,
once one introduces the auxiliary field $b_\mu$, a BRST 
invariance of the kind arising in topological
field theories \cite{BS}-\cite{BRT} can be unraveled.
The use of BRST technique for bosonization of fermion models
was initiated in the developement of the smooth bosonization
approach \cite{DNS1}-\cite{DNS3},
\cite{DNS4}-\cite{DNS5}, 
closely related to bosonization duality and 
to the present
treatment. In the present case, it
 allows to factor out the auxiliary field
in the same way Polyakov-Wiegmann identity did
the job in $d=2$.

The resulting  bosonization  action coincides with that obtained using
a completely different approach \cite{B},
 based in the use of an interpolating Lagrangian \cite{VN}-\cite{vN}.
The advantage of the present method lies in the fact that the BRST
symmetry can be formulated in arbitrary dimensions while the
interpolating Lagrangian, which replaces the role of this
symmetry in decoupling
 auxiliary and bosonic fields  is in principle applicable only in odd-dimensional spaces.

We  consider $N$ massive Dirac fermions in $d=3$ Euclidean dimensions
with Lagrangian
\beq
L = \bar\psi (\id   + m ) \psi
\label{3L}
\eeq
The corresponding generating functional reads
\beq
Z_{fer}[s] =  \int \D \bar \psi \D \psi
\exp[-\int d^3x \bar \psi (\id + \qs + m ) \psi ]
\label{41}
\eeq
Again, we introduce an auxiliary  vector field
$b_\mu$ and use the $d=3$ identity
(proven in the Appendix)
\beq
Z_{fer}[s] = X[s]^{-1}\!\!
\int \!\!Db_\mu X[b]
\det(2\varepsilon_{\mu \nu \alpha} D_\nu[b])
 \delta( {^*\!\!f}_\mu[b] - {^*\!\!f}_\mu[s] )
\det(\id + m + \bs) \label{IX}
\eeq
Here
\beq
^*\!\!f_\mu = \varepsilon_{\mu \nu \alpha}
f_{\nu\alpha}
\label{II}
\eeq
Concerning $X[b]$, it is an arbitrary functional which can be introduced
in order to control the issue of symmetries at each stage of our
derivation.
Indeed, bosonization
of three dimensional very massive fermions ends with a
bosonic field with dynamics governed by a
Chern-Simons action.  As explained in \cite{FAS}, an 
appropriate choice 
of $X$ allows to end with the natural gauge connection
transformation law for this bosonic field. Following
 \cite{FAS}, we choose $X$ in the form
\beq
X[b] = \exp(\mp \frac{i}{24\pi} \varepsilon_{\mu\nu\alpha} 
tr \int d^3x b_\mu b_\nu b_\alpha )
\label{x}
\eeq
We can see at this point how an exact bosonization rule for the fermion
current can be derived independently of the fact that one cannot
calculate exactly the fermion determinant for $d > 2$.
Indeed, if  we introduce a Lagrange
multiplier $A_\mu$ to represent the delta function, we can write $Z_{fer}$
in the form
\beq
Z_{fer}[s] = X[s]^{-1}
\int \D A_\mu \exp \left( \mp \frac{i}{16\pi}
tr \int d^3x  A_\mu  {^*\!\!f}_\mu[s] \right) \times
\exp(-S_{bos}[A]) \label{ches}
\eeq
where we have defined the bosonic action $S_{bos}[A]$ 
as
\begin{eqnarray}
\exp(-S_{bos}[A]) & = & \int \D b_\mu \det(\id + m + \bs) X[b] \times
\nonumber \\
& & \det(2\varepsilon_{\mu \nu \alpha}D_\nu[b])
\exp\left( \pm \frac{i}{16\pi}
tr \int d^3x  A_\mu  {^*\!\!f}_\mu[b] \right)
\label{jeje}
\end{eqnarray}
With the choice (\ref{x}) one indeed has gauge invariance
of $S_{bos}[A]$, $A_\mu$ and $b_\mu$ both transforming as gauge fields,
and one also explicitely verifies eq.(\ref{idiss}).

Then, from eq.(\ref{ches}) we have
\beq
\bar \psi \gamma_\mu t^a \psi \to
\pm \frac{i}{8\pi}\varepsilon_{\mu \nu \alpha}
\partial_\nu A^a_\alpha \,\, .
\label{Xsss}
\eeq
In writing eq.(\ref{Xsss}) we have ignored terms quadratic
and cubic in the source
which, as in $d=2$, are irrelevant for the current algebra.
Correlation functions of currents pick a contribution 
from these terms, as already discussed
in other approaches to bosonization \cite{dVR}-\cite{F}. Having
these terms local support, they do not contribute
to the current commutator algebra. (That this is so
can be easily seen using for example the 
Bjorken-Johnson-Low method).

We insist that our result (\ref{Xsss}) does not imply any kind
of approximation. However, to achieve 
a complete bosonization, one needs an explicit local
form for the bosonic
action and it is at this point where
approximations have to be envisaged so
as to evaluate the fermion determinant. In $d=3$ dimensions 
this determinant cannot be
computed exactly. However, all approximation approaches and
regularization schemes have shown the occurrence
of a parity violating Chern-Simons term together with
parity conserving terms which can be computed approximately.
We shall use the result obtained by making an expansion
in inverse powers of the fermion mass \cite{Red},
\beq
\ln \det (\id + m + \bs) =
  \pm \frac{i}{16\pi} S_{CS}[b] +
  I_{PC}[b] +
  O(\partial^2/m^2)  ,
\label{9f}
\eeq
where the Chern-Simons action $S_{CS}$ is given by
\beq
  S_{CS}[b] = \int\dv
 \varepsilon_{\mu\nu\lambda} \tr \int\dv
 (
   f_{\mu \nu} b_{\lambda} -
   \frac{2}{3} b_{\mu}b_{\nu}b_{\lambda}
  )  .
\eeq
Concerning the parity conserving contributions, one has
\beq
I_{PC}[b] =
  - \frac{1}{24\pi  m} \tr\int\dv f^{\mu\nu} f_{\mu\nu}
  + \cdots  ,
\label{8f}
\eeq
We can then write, up to corrections of order $1/m$, the bosonic
action $S_{bos}[A]$
 in the form
(From here on we shall omit to indicate the trace {\it tr}
for notation simplicity)
\begin{eqnarray}
\exp(-S_{bos}[A]) & = & \int \D b_\mu \, X[b] 
 \exp(\pm \frac{i}{16\pi} S_{CS}[b]) \times
\nonumber \\
& & \det(2\varepsilon_{\mu \nu \alpha}D_\nu[b])
\exp\left( \pm \frac{i}{16\pi}
\int d^3x  A_\mu  {^*\!\!f}_\mu[b] \right)
\label{ufin}
\end{eqnarray}

We shall now introduce ghost fields ${\bar c}_\alpha$ and
$c_\alpha$ to write the determinant in the r.h.s. of eq.(\ref{ufin}).
With this, $Z_{fer}[s]$ takes the form
\begin{eqnarray}
Z_{fer}[s] & = & X[s]^{-1}
\int \D b_\mu \D {\bar c}_\alpha  \D c_\alpha \D A_\mu
\exp \left( \mp\frac{i}{16\pi}
\int d^3x  A_\mu  {^*\!\!f}_\mu[s] \right) \times
\nonumber \\
& & \exp(-S_{eff}[b,A,\bar c, c]) 
\label{312}
\end{eqnarray}
with
\begin{eqnarray}
S_{eff}[b,A,\bar c, c] & = &
  \mp \frac{i}{16\pi} S[b] \nonumber \\
& &   \mp \frac{i}{8\pi} \varepsilon_{\mu \nu \alpha} \int  d^3x
( A_\mu (\partial_\nu b_\alpha + b_\nu b_\alpha)
-  {\bar c}_\mu  D_\nu [b] c_\alpha 
)
\label{312S}
\end{eqnarray} 
and
\beq
S[b] = 2  \varepsilon_{\mu \nu \alpha} \int d^3x
b_\mu(\partial_\nu b_\alpha +\frac{1}{3} b_\nu b_\alpha)
\label{sssx}
\eeq
At this point we have arrived to an exact bosonization recipe for the fermion
current, eq.(\ref{Xsss}), but we still need
an explicit formula for the
bosonic action as a functional of $A_\mu$. This requires integration over
the auxiliary
fields $b_\mu$, $\bar c_\mu$ and $c_\mu$ of the complicated effective action
$S_{eff}$ as defined by eq.(\ref{312S}). In the two-dimensional case, this
last step was possible because Polyakov-Wiegmann identity allowed us to
decouple the auxiliary fields from the bosonic field ($A_\mu$). In
the present case, integration will be possible because of the existence
of an underlying BRST invariance that can be made apparent in $S_{eff}$.
In order to directly get an {\it off-shell} nilpotent set of BRST 
transformations leaving invariant the effective action, we shall
introduce additional auxiliary fields \cite{bas}, thus writing
\begin{eqnarray}
 Z_{fer}[s] & = & X[s]^{-1}
\int \D b_\mu \D {\bar c}_\alpha  \D c_\alpha \D A_\mu \D h_\mu \D l
\D \bar \chi
\nonumber \\
& & \exp \left( \mp\frac{i}{16\pi}
\int d^3x  A_\mu  {^*\!\!f}_\mu[s] \right) 
 \exp(-{\tilde S}_{eff}[b,A,\bar c, c, h, l, \bar \chi]) 
\label{barra}
\end{eqnarray}
with  $\tilde S_{eff}$ defined as
\begin{eqnarray}
& & {\tilde S}_{eff}[b,A,\bar c, c, h, l, \chi] = 
   \mp \frac{i}{16\pi} S[b-h]  
 \mp \frac{i}{16\pi} \int d^3x 
(l h_\mu h_\mu - 2 \bar \chi h_\mu c_\mu ) \nonumber \\
& &   \mp \frac{i}{8\pi} \varepsilon_{\mu \nu \alpha} \int  d^3x
( A_\mu (\partial_\nu b_\alpha + b_\nu b_\alpha)
-  {\bar c}_\mu  D_\nu [b] c_\alpha 
 )
\label{312SS}
\end{eqnarray}
Integration over the auxiliary field $l$ makes $h_\mu = 0$ 
this showing the equivalence of eq.(\ref{barra}) and eq.(\ref{312}).
Now, the effective action ${\tilde S}_{eff}$ is invariant
under BRST transformations defined as
\[
\delta {\bar c}_\alpha = A_\alpha 
\;\;\;\;\;\;
\delta A_\alpha = 0
\]
\[
\delta b_\alpha = c_\alpha 
\;\;\;\;\;\;
\delta c_\alpha = 0
\]
\beq
\delta h_\alpha = c_\alpha 
\;\;\;\;
\delta \bar \chi = l
\;\;\;\;
\delta l = 0
\label{trb}
\eeq
This
BRST transformations are
 related to those employed
in the smooth bosonization \cite{DNS1}-\cite{DNS3}, 
\cite{DNS4}-\cite{DNS5} approach and
resemblant of those arising
in topological field theories \cite{BS}-\cite{BRT}. For example, in
$d=4$ topological
Yang-Mills theory the invariance of the starting classical action
(the Chern-Pon\-trya\-gin topological charge)
 under the most general transformation of the gauge field,
$b_\mu \to b_\mu + \epsilon_\mu$, leads to a BRST
transformation for $b_\mu$ of the form $\delta b_\mu = c_\mu$, which
corresponds to that in formula (\ref{trb})  \cite{Wi0}-\cite{BS}.
Closer to our model are the so-called
Schwartz type topological theories which
include the Chern-Simons theory and the
BF model
analyzed in detail in refs.\cite{hor}-\cite{BRT}. It should be
stressed that the topological character of the
effective action (\ref{312S}) exclusively concerns
the large fermion mass regime where the fermion determinant can be
written in terms of the CS action.

Now, using transformations (\ref{trb}), $\tilde S_{eff}$
can be compactly written  in the form
\beq
\tilde S_{eff}[b,A,\bar c, c] =
\mp  \frac{i}{16\pi} S[b-h] \mp
\frac{i}{8\pi} \int d^3x \, \delta{\cal F}[\bar c,b,h,\bar \chi]
\label{sss}
\eeq
with
\beq
 {\cal F} = \varepsilon_{\mu \nu \alpha}
  \bar c_\mu(\partial_\nu b_\alpha + b_\nu b_\alpha)
+ \frac{1}{2} \bar \chi h_\mu h_\mu 
\label{off}
\eeq
At this point,  an arbitrary functional ${\cal G}$
may be added to ${\cal F}$ without changing
the partition function since it will  enter
in $Z_{fer}$  as an exact BRST form. The
idea is to choose  ${\cal G}$  so as to decouple
the auxiliary field $b_\mu$ (to be integrated out
afterwards) from the vector field $A_\mu$
which will be the bosonic counterpart of the original
fermion field.
We shall then consider
\beq
{\cal F} \to {\cal F} + {\cal G}
\label{s1}
\eeq
with
\begin{eqnarray}
{\cal G} & = &
\frac{1}{2} \varepsilon _{\mu \nu \alpha }\bar c_\mu \;
([b_\nu,A_\alpha] + [A_\nu,A_\alpha] +C[b_\nu,h_\alpha]
+(1+C)[A_\nu,h_\alpha]  \nonumber \\
& &  - (C+1) [h_\nu,h_\alpha] +2\partial_\nu b_\alpha
+4\partial_\nu A_\alpha +2C \partial_\nu h_\alpha  )
\label{el1}
\end{eqnarray}
Here $C$ is an arbitrary constant.

The addition of $\delta {\cal G}$ allows us to make
contact at this point with the effective action discussed in
refs.\cite{B},\cite{vN}. Indeed,
after the shift 
\beq
 b_\mu \to 2b_\mu - A_\mu + h_\mu
\label{LG}
\eeq
(the new $b_\mu$ transforms again as
a gauge connection, with $h_\mu$ transforming covariantly)
the Lagrange multiplier $A_\mu$  
(which will play the role of the bosonic field in our bosonization
approach, as identified by the
source term)
completely decouples for $h_\mu = 0$, so that
integrating out auxiliary fields we end with
\beq
Z_{fer}[s] = {\cal N}X[s]^{-1} 
\int \D A_\mu
\exp ( \mp\frac{i}{16\pi}
\int d^3x  A_\mu  {^*\!\!f}_\mu[s] ) \times
 \exp(\pm \frac{i}{16\pi}S_{CS}[A]).
\label{chesi}
\eeq
Here ${\cal N}$ is a constant (i.e. it 
is independent of the source) resulting from integration
of auxiliary, ghosts and the $b$ field,
\beq
{\cal N} = \int \D b_\mu \,\, 
det\left( 2(2+C)\varepsilon_{\mu\nu\alpha} D_\nu[b] \right)
\exp(\pm \frac{i}{4\pi} S_{CS}[b]) 
\label{final}
\eeq

We have then the bosonization result
\beq
Z_{fer} [s] \approx Z_{CS}[s]
\label{uli}
\eeq
where $\approx$ means that our result is valid up to $1/m$ corrections
since we used a result for the fermion
determinant which is valid up to this order. 
We then see that we have ended with a Chern-Simons action as the bosonic
equivalent of the original free fermion action with a coupling
to the external source $s_\mu$  of the form $ A_\mu  {^*\!\!f}_\mu[s] $.

In considering fermion current bosonization 
within the $1/m$ approximation, the following facts 
should be taken into account.
It is at the lowest order in $1/m$ that
the resulting bosonic action is topological and a
 large BRST invariance is unraveled.
Now, using
the freedom to
modify the action by BRST exact forms, one 
could think of adding to the topological bosonic action
 terms of the form $ \delta{\cal H}$ with
\beq
{\cal H} = \int d^3x \  \varepsilon_{\mu \nu \alpha}
{\bar c}_{\mu} {\cal H}_{\nu \alpha}[s] \, ,
\label{ad1}
\eeq
with ${\cal H}_{\mu \nu}[s]$  an arbitrary functional of the external
source $s_{\mu}$.  In particular, choosing adequately ${\cal H}$ one
could think of changing or even 
eliminating, to this order in $1/m$, the source dependence from 
$Z_{fer}[s]$.  
Now, this is a characteristic of
Schwarz like topological models \cite{BRT}. In particular, 
the phase space of the Chern-Simons theory is
the moduli space of flat connections on the given space manifold.
So, if one looks {\it up to this order in $1/m$},
at the
 generating functional of current Green
functions, one has, from Eqs.(\ref{41})  and (\ref{9f}) that the generating functional of connected Green functions is
precisely a Chern-Simons action for the source $s$,
\beq
W[s]=- log Z[s] = \mp \frac{i}{16\pi}  S_{CS}[s].
\label{ad3}
\eeq
Thus, making functional derivatives in the above expression with respect
to the source and then putting the sources to zero, all
current vacuum expectation values vanish identically up to contact terms
(these terms, derivatives of the Dirac delta function, also vanish if we
regularize appropriately the product of operators at coincident points.
Moreover our results are valid in the $m\to\infty$ limit where the deep
ultraviolet region is excluded). 
Non-vanishing observables are in fact
topological objets, non local functionals
of $A_\mu$ (Wilson loops) that
are in correspondance to knots polynomial invariants. 
Hence the bosonization recipe
(\ref{Xsss}) when used
to this order in $1/m$
 makes sense if one is to calculate vacuum expectation 
values of fermion
objects leading for example to holonomies in terms of the bosonic
field $A_\mu$. This calculation was discussed at length in \cite{B}.

\section{Summary}

We have shown in this paper that the path-integral bosonization
approach   developed in previous investigations
\cite{FS}-\cite{B} is well-suited to study fermion models
in $d \ge 2$ dimensions when a non-Abelian symmetry
is present.

We have started by reobtaining
in Section 2 the well-honoured non-Abelian bosonization
recipe for   two dimensional
massless fermions. Although
well-known, this result allowed us to identify the point
in which the non-Abelian character of the symmetry
makes difficult the factorization of the path-integral
which will represent the partition function of the
resulting bosonic model. In two dimensions this
factorization can be seen as a result of the existence
of the Polyakov-Wiegmann identity for Wess-Zumino-Witten actions,
and this can {\it a priori} put some doubts on the possibility of 
extending
the approach to $d>2$.

That also in $d=3$ one can obtain very simple bosonization
rules for the non-Abelian case is the main result of
section 3. 
Concerning the fermion current, we obtained an
exact bosonization result which is the
natural extension of the Abelian case.
In respect with the bosonization recipe for the fermion action,
we considered
the case of very massive fermions
for which the fermion determinant
is related to the non-Abelian Chern-Simons action.
In this case the factorization of the auxiliary
and Lagrange multiplier fields is achieved after
discovering a BRST invariance reminiscent of that at the
root of topological models and
related to that exploited in the
smooth bosonization approach \cite{DNS1}-\cite{DNS3}, 
\cite{DNS4}-\cite{DNS5}.
 Addition of BRST exact terms
allows us  to extract the partition function for the
boson counterpart of the original fermion fields.

Our bosonization method starts by introducing
in the fermionic generating functional  an 
auxiliary field as it is done in the 
smooth bosonization and duality
approaches to bosonization \cite{DNS1}-\cite{mar}. It becomes 
clear in our approach that, for non-Abelian
symmetries, it is crucial to include the
``Faddeev-Popov'' like determinant which
accompanies the delta function imposing a condition
on the auxiliary field curvature. In fact, the BRST symmetry
which allowed to arrive to the correct bosonic generating functional
can be seen as a result of this fact and related to
the way in which BRST symmetry can be unraveled
by a change of variables as advocated in ref.\cite{bas}.

It should be stressed that the only approximation in our
approach stems from the
necessity of evaluating the fermion determinant which, in
$d > 2$, implies  some kind of expansion. In the
present work we have used a result valid for very massive
fermions but one can envisage  approximations which
can cover other regimes, in particular the massless case.
This was considered for the abelian case in \cite{BFO} and
the corresponding  bosonization  analysis  thoroughly
discussed in \cite {LNS}. We expect that a similar analysis
can be done in the non-abelian case and we hope to report
on it in a future paper.


\newpage

\section*{Appendix}

\subsection*{ $d=2$}

We shall  prove  identity (\ref{idle}) used in our derivation 
of $d=2$ bosonization rules,
\beq
 \int \D b_\mu {\cal H}[b]
\delta\left[\varepsilon_{\mu \nu} (f_{\mu\nu}[b] - f_{\mu\nu}[s])\right]
=\int \D g {\cal H}[s^g]
\label{idlex}
\eeq
where ${\cal H}$ is a gauge-invariant functional. Note
that in eq.(\ref{idlex}) it is implicit that $b_\mu$ should
be treated as a gauge field and hence a gauge fixing is
required.
A convenient gauge choice
is
\beq
b_+ = s_+
\label{otrav}
\eeq
so that  identity (\ref{idlex}) takes the form
\beq
 \int \D b_+ \D b_- \;\Delta \delta(b_+ - s_+){\cal H}[b_+,b_-]
\delta[\varepsilon_{\mu \nu} (f_{\mu\nu}[b] - f_{\mu\nu}[s]]
 = \int \D g {\cal H}[s^g]
\label{idleg}
\eeq
with $\Delta$ the Faddeev-Popov determinant for gauge condition
(\ref{otrav}),
\beq
\Delta = det D^{Adj}_+[s_+]
\label{FP}
\eeq
We now prove eq.(\ref{idleg}).
Let us start from the l.h.s. of eq.(\ref{idleg})
performing first the $b_+$ trivial integration and then the
$b_-$ one
\begin{eqnarray}
& &  \int \D b_+ \D b_- \Delta \delta(b_+ - s_+){\cal H}[b_+,b_-]
\delta(\varepsilon_{\mu \nu} (f_{\mu\nu}[b] - f_{\mu\nu}[s]) = \nonumber\\
& &
\Delta[s_+] \int \D b_- {\cal H}[s_+,b_-]
\delta( D_+[s_+] b_- - D_+[s_+]s_- ) = \nonumber\\
& &  \frac{\Delta[s_+]}{det D^{Adj}_+[s_+]} \int \D b_- {\cal H}[s_+,b_-]
\delta(b_- - s_-)
 =  {\cal H}[s_+,s_-]
\label{largui}
\end{eqnarray}
In the last line we have used the explicit form of the Faddeev-Jacobian to
cancel out both determinants.
Being ${\cal H}[s_+,s_-]$ gauge independent, we can rewrite
(\ref{largui}) in the form (appart from a gauge group volume factor)
\beq
 \int \D b_+ \D b_- \Delta \delta(b_+ - s_+){\cal H}[b_+,b_-]
\delta \left[\varepsilon_{\mu \nu} (f_{\mu\nu}[b] - f_{\mu\nu}[s])
\right] =
\int \D g {\cal H}[s_+^g,s_-^g]
\label{uno}
\eeq
Identity (\ref{idlex}) is then proven.

\subsection*{ $d=3$}

The 
proof of identity (\ref{IX}), the analogous in the $d=3$ case of
(\ref{idlex}),  is very simple. One 
wishes to prove that
the generating functional $Z_{fer}[s]$ in the presence of a source
$s_\mu$,
\beq
Z_{fer}[s] = \det (\id + m + \qs) = F[s]
\label{leg16}
\eeq
can be written in the form
\beq
Z_{fer}[s] = {X[s]}^{-1} \int \D b_\mu \,  X[b]F[b] 
\,
\det( 2\varepsilon_{\mu \nu \alpha} D_\nu[b] ) \,
 \delta\left( \varepsilon_{\mu \nu \alpha} (f_{\nu \alpha}[b]
- f_{\nu \alpha}[s])\right)
\label{leg17}
\eeq
Here $X[b]$ is an arbitrary functional of $b_\mu$ satisfying
$X[0^g]=1$. As advocated in \cite{FAS}, its introduction
allows to end with a  model in which the bosonic field transforms
as a connection, this being consistent with the fact
its dynamics is governed by a Chern-Simons action.

The proof of eq.(\ref{leg17}) is based on the
well-known identity 
\beq
\delta(H[b]) = [\det (\delta H/\delta b)]^{-1}  \delta(b - b^*)
\eeq
with $H[b^*] = 0$ and the fact that
the equation
$f_{\mu\nu}[b] = f_{\mu\nu}[s]$
has the unique solution $ b_\mu = s_\mu$.

Let us end by noting that if one compares  formula (\ref{leg17})
in $d=3$ dimensions with the corresponding one in $d=2$ (for example
the identity  (\ref{idlex})), one sees that a determinant equivalent to that
appearing in the former is absent in the latter. This is due
to the fact that the curvature condition requires three delta functions
in $d=3$ dimensions but only  one in $d=2$. Handling these
delta functions leaves a jacobian
in three dimensions while no jacobian remains in two dimensions.

\vspace{1 cm}

\underline{Acknowledgements}: The authors wish to thank Matthias Blau,
Daniel Cabra, Fran\c{c}ois Delduc, C\'esar Fosco, Eric Ragoucy
and Frank Thuillier for helpful discussions
and comments. F.A.S. and
C.N.  are partially suported
by Fundacion Antorchas, Argentina and a
Commission of the European Communities
contract No:C11*-CT93-0315.


\begin{thebibliography}{99}
%
\bibitem{FS} E.~Fradkin and F.A.~Schaposnik,
Phys. Lett. {\bf B338} (1994) 253.
\bibitem{FAS} F.A.~Schaposnik, Phys. Lett. {\bf B356} (1995) 39.
\bibitem{LNS} J.C.~Le Guillou, C.N\'u\~nez and F.A.~Schaposnik,
Ann. of Phys. (N.Y.) in press.
\bibitem{B} N.~Brali\'c, E.~Fradkin, M.V.~Man\'\i as
and F.A.~Schaposnik, Nucl.Phys. {\bf B446} (1995) 144.
\bibitem{DNS1} P.H.~Damgaard, H.B.~Nielsen and R.~Sollacher,
Nucl. Phys. {\bf B385} (1992) 227.
\bibitem{DNS2} P.H.~Damgaard, H.B.~Nielsen and R.~Sollacher,
Phys. Lett.  {\bf B296} (1992) 132.
\bibitem{DNS3} P.H.~Damgaard, H.B.~Nielsen and R.~Sollacher,
Nucl. Phys. {\bf B414} (1994) 541.
\bibitem{BQ}  C.P.~Burgess and F.~Quevedo, Nucl. Phys. {\bf B421} (1994) 373.
\bibitem{BLQ}  C.P.~Burgess an, C.A.~L\"utken and F.~Quevedo,
Phys. Lett. {\bf B326} (1994) 18.
\bibitem{BQn} C.P.~Burgess and F.~Quevedo, Phys. Lett. {\bf B329} (1994) 457.
\bibitem{DNS4} P.H.~Damgaard and R.~Sollacher,
Nucl. Phys. {\bf B433} (1995) 671.
\bibitem{DNS5} P.H.~Damgaard, F.~De Jonghe and R.~Sollacher,
Nucl. Phys. {\bf B454} (1995) 701.
\bibitem{TSSG} A.N.~Theron, F.A.~Schaposnik, F.G.~Scholtz and
H.B.~Geyer, Nucl.Phys. {\bf B437} (1995) 187.
\bibitem{CRV} J.L.~Cort\'es, E.~Rivas and L.~Vel\'azquez,
Phys. Rev. {\bf D53} (1996) 5952.
\bibitem{mar} P.A. Marchetti in Common Trends in
Condensed Matter and High Enery Physics, Chia Laguna, Italy,
Sep 1995. hep-th/9511100 report.
\bibitem{lut} A. Luther, Phys. Rev. {\bf D19} (1979) 320.
\bibitem{lusch} M.~Luscher, Nucl. Phys. {\bf B326} (1989) 557.
\bibitem{hal} F.D.M. Haldane, Helv. Phys. Acta {\bf 65} (1992) 52.
\bibitem{kov} A. Kovner and P.S. Kurzepa, Phys. Lett. {\bf B321} (1994) 129.
\bibitem{Mar}E.C.~Marino, {\it Phys. Lett.} {\bf B263} (1991), 63.
\bibitem{Fro}  J. Frohlich, R. G\"otschmann and P.A. Marchetti,
J.Phys. {\bf A28} (1995) 1169.
\bibitem{BBG} N.~Banerjee, S.~Banerjee and S.~Gosh,  hep-th/9607065,
unpublished report.
\bibitem{BFO} D.G.~Barci, C.D.~Fosco and L.E.~Oxman,
Phys. Lett. {\bf B375} (1996) 267.
\bibitem{PWI} A.M.~Polyakov and P.B.~Wiegmann, Phys. Lett. {\bf 131B}
(1983) 121; Phys Lett. {\bf 141B} (1984) 223.
\bibitem{Wi} E.~Witten, Commun. Math. Phys. {\bf 92} (1984) 455.
\bibitem{Wi0} E.~Witten, Commun. Math. Phys. {\bf 117} (1988) 353.
\bibitem{BS} L.~Baulieu and I.M.~Singer, Nucl. Phys. (Proc. Suppl.)
{\bf 5B} (1988) 12.
\bibitem{hor} G.~Horowitz, Comm. Math. Phys. {\bf 125} (1989) 417.
\bibitem{BRT} D.~Birmingham, M.~Blau,
M.~Rakowski and G.~Thompson, Phys. Rep. {\bf 209}
(1991) 129.
\bibitem{dVR} P.~di Vecchia and P.~Rossi, Phys. Lett. {\bf 140B } (1984) 344.
\bibitem{F} Y.~Frishmann, Phys. Lett. {\bf 146B} (1984) 204.
\bibitem{GO} P.~Goddard and D.~Olive,
Int. Jour. Mod. Phys. {\bf A1} (1986) 303.
\bibitem{FNS} E.~Fradkin, C.M.~Na\'on and F.A.~Schaposnik,
Phys. Rev. {\bf D36} (1987) 3809.
\bibitem{dVDP} P.~di Vecchia, B.~Durhus and J.L.~Petersen, Phys. Lett.
{\bf 144B} (1984) 245.
\bibitem{GR} D.~Gonz\'ales and A.N.~Redlich, Phys. Lett. 147B (1984) 150.
\bibitem{NS} A.T.~Niemi and G.W.~Semenoff, Phys. Rev.
Lett. {\bf 51} (1983) 2077
\bibitem{Red} A.N.~Redlich, Phys. Rev. Lett. {\bf 52} (1984) 18;
Phys. Rev. {\bf D29} (1984)236.
\bibitem{Wit} E.~Witten, Commun. Math. Phys. {\bf 144} (1992) 189.
\bibitem{VN} P.K.~Townsend, K.~Pilch and
P.~van~Nieuwenhuizen, Phys.~Lett. {\bf B136} (1984) 38; {\bf B137} (1984) 443.
\bibitem{DJ} S.~Deser and R.~Jackiw, Phys.Lett. {\bf B139} (1984) 371.
\bibitem{vN} A.~Karlhede, U.~Lindstr\"om, M.~Ro\v{c}ek and
  P.~van~Nieuwenhuizen, Phys.~Lett. {\bf 186B} (1987) 96.
\bibitem{bas} F.~Bastianelli, Nucl. Phys. {\bf B361} (1991) 555.
%
\end{thebibliography}
\end{document}